\documentclass{emulateapj}
\usepackage{apjfonts}

\shorttitle{{\em WMAP} 3 year data and Reionization}
\shortauthors{Alvarez et al.}

\def\plotone#1{\centering \leavevmode
\epsfxsize= 1.0\columnwidth \epsfbox{#1}}

\begin{document}
\title{
Implications of {\em WMAP} 3 Year Data for the Sources of Reionization
}

\author{
Marcelo A. Alvarez\altaffilmark{1}, 
Paul R. Shapiro\altaffilmark{1}, 
Kyungjin Ahn\altaffilmark{1},
and Ilian T. Iliev\altaffilmark{2}
}
\submitted{Accepted to ApJ Letters}

\altaffiltext{1}{
  Department of Astronomy, University of Texas at Austin,
  1 University Station, C1400, Austin, TX 78712
}
\altaffiltext{2}{
Canadian Institute for Theoretical Astrophysics, University of
  Toronto, 60 St. George Street, Toronto, ON M5S 3H8, Canada
}

\begin{abstract}
New results on the anisotropy of the cosmic microwave background (CMB)
and its polarization based upon the first three years of data from the
{\em Wilkinson Microwave Anisotropy Probe} ({\em WMAP}) have revised the
electron scattering optical depth downward from $\tau_{\rm es}=
0.17^{+0.08}_{-0.07}$ to $\tau_{\rm es}=0.09\pm 0.03$. This implies
a shift of the effective reionization redshift from $z_r\simeq 17$ to 
$z_r\simeq 11$.  Previous attempts to explain the high 
redshift of reionization inferred from the {\em WMAP} 1-year
data have led 
to widespread speculation that the sources of reionization must have 
been much more efficient than those associated with the star formation 
observed at low redshift. This is consistent, for example, with the 
suggestion that early star formation involved massive, Pop~III stars 
which early-on produced most of the ionizing radiation escaping from 
halos. It is, therefore, tempting to interpret the new {\em WMAP} results as 
implying that we can now relax those previous high demands on the 
efficiency of the sources of reionization and perhaps even turn the 
argument around as evidence {\em against} such high efficiency. We show 
that this is not the case, however.  The new {\em WMAP} results also find 
that the primordial density fluctuation power spectrum has a lower 
amplitude, $\sigma_8$, and departs substantially from the 
scale-invariant spectrum. We show that these effects combine to cancel 
the impact of the later reionization implied by the new value of 
$\tau_{\rm es}$ on the required ionizing efficiency per collapsed baryon.  
The delay of reionization is surprisingly well-matched by a comparable 
delay (by a factor of $\sim 1.4$ in scale factor) in the formation of 
the halos responsible for reionization. 
\end{abstract}


\section{Introduction}\label{sec:introduction}
One of the most important outstanding problems in cosmological
structure formation is how and when the universe was reionized.  
Observational
constraints such as the Thomson scattering optical depth to the last
scattering surface (Kogut et al. 2003; Page et al. 2006) from the 
large-angle polarization anisotropy in the CMB detected by {\em WMAP}
and the intergalactic, hydrogen Ly$\alpha$ absorption spectra of high-redshift
quasars (e.g., Becker et al. 2001)
provide crucial constraints on the theory of cosmic reionization and
the structure formation which caused it
during the early epochs that have thus far escaped
direct observation.
The {\em WMAP} first-year data implied an electron scattering optical
depth, $\tau_{\rm es}=0.17$, which seemed surprisingly large at the
time, since it was well in excess of the value, $\tau_{\rm es}\simeq 0.04$, for
an intergalactic medium (IGM) abrubtly ionized at $z_r\simeq 6.5$, the
reionization epoch which had been suggested by quasar measurements of
the Gunn-Peterson (Gunn \& Peterson 1965; ``GP'') effect.
In order for such an abrupt reionization to explain the
high value of 0.17 observed by {\em WMAP} for $\tau_{\rm es}$, in fact,
$z_r\simeq 17$ is required.  This presented a puzzle for the theory of
reionization: How was reionization so advanced, so {\em early} in our
observed $\Lambda$CDM universe, and yet so {\em extended} in time as to
accumulate the high $\tau_{\rm es}$ observed by WMAP, while {\em
ending} as late as $z\simeq 6.5$ to satisfy the quasar spectral constraints?

This stimulated widespread speculation regarding the efficiency for
the formation of the early stars and/or miniquasars which were the
sources of reionization, as well as for the escape of their ionizing
photons into the IGM (e.g., Haiman \& Holder 2003; Cen 2003; Wyithe
\& Loeb 2003; Kaplinghat et al. 2003; Sokasian et al. 2004;   Ciardi,
Ferrara, \& White 2003; Ricotti \& Ostriker 2004).  A general
consensus emerged that the 
efficiencies for photon production and escape associated with
present-day star formation were not adequate to explain the early
reionization implied by the high $\tau_{\rm es}$ value, given the rate
of early structure formation expected in the $\Lambda$CDM universe.
Common to most attempts to explain the high $\tau_{\rm es}$ was the
assumption that early star formation favored massive Population III
stars, either in ``minihalos,'' with virial temperatures $T_{\rm vir}<
10^4$ K, requiring that $H_2$ molecules cool the gas to enable star
formation (Abel, Bryan, \& Norman 2002; Bromm, Coppi, \& Larson 2002),
or else in larger halos with $T_{\rm vir}>10^4$~K, for which atomic
hydrogen cooling is possible, instead.  A high efficiency for turning
halo baryons into stars and a high escape fraction for the ionizing
radiation into the IGM were generally required as well.  Several
effects were suggested that could extend the reionization epoch,
too, including the rising impact of small-scale structure as a sink of
ionizing
photons (e.g., Shapiro, Iliev, \& Raga 2004; Iliev, Shapiro, \&
Raga 2005; Iliev, Scannapieco, \& Shapiro 2005), the suppression of
low-mass source-halo formation inside the growing intergalactic H~II
regions (e.g., Haiman \& Holder 2003), and a general decline of the
efficiency for releasing ionizing radiation over time (e.g., Cen 2003;
Choudhury \& Ferrara 2005).

With three years of polarization data, {\em WMAP} (henceforth,
``WMAP3''), has now produced a more accurate determination of
$\tau_{\rm es}$, which revises the optical depth downward to
$\tau_{\rm es}=0.09\pm 0.03$ (Page et al. 2006).  This value is
consistent with an abrupt reionization at $z_r=11$, significantly
later than that implied by the {\em WMAP} first-year data (henceforth,
``WMAP1''').  It is natural to wonder if this implies that the high
efficiency demanded of ionizing photon production by WMAP1, described
above, can now be reduced, accordingly, to accommodate the later epoch
of reionization determined by WMAP3.  In what follows, we will show
that this is not the case.

\begin{figure}[t]
\plotone{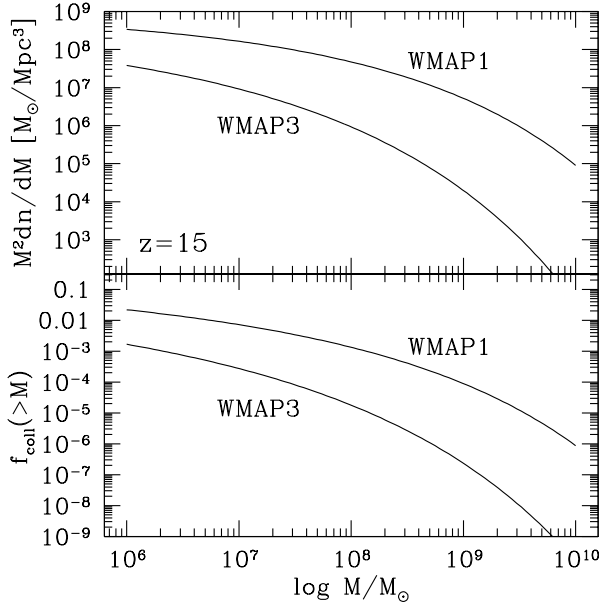}
\caption{Halo abundance vs. mass for new (WMAP3) and old
(WMAP1) parameters at $z=15$, as labelled.  {\em Top:}
Press-Schechter mass function. {\em Bottom:} Fraction of matter
$f_{\rm coll}(>M)$ in halos with mass greater than $M$.  }
\vspace{-0.7cm}
\end{figure}

Structure formation in the $\Lambda$CDM universe with the primordial density
fluctuation power spectrum measured by WMAP3 is delayed relative to
that in the WMAP1 universe, especially on the small-scales responsible
for the sources of reionization.  This, by itself, is not surprising,
since there was always a degeneracy inherent in measuring the
amplitude of the primordial density fluctuations using the CMB
temperature anisotropy alone, resulting from the unknown value of
$\tau_{\rm es}$.  Higher values of $\tau_{\rm es}$, that is, imply
higher amplitude density fluctuations to produce the same level of CMB
anisotropy. This degeneracy is broken by the independent measurement
of $\tau_{\rm es}$ made possible by detecting the polarization
anisotropy, as well. Hence, when WMAP3 revised the value of $\tau_{\rm
es}$ downward relative to WMAP1, so it revised downward the amplitude
of the density fluctuations.  This same decrease of $\tau_{\rm es}$
implies a tilt away from the scale-invariant power spectrum,
$P(k)\propto k^{n_s}$ with $n_s=1$, which lowers the
density fluctuation amplitude on small scales more than on large
scales.  As we will show, this delays the structure formation which
controls reionization by just the right amount such that, if
reionization efficiencies were large enough to make reionization
early and $\tau_{\rm es}=0.17$ in the WMAP1 universe, the same
efficiencies will cause reionization to be later in the WMAP3 universe
and $\tau_{\rm es}\sim 0.09$, as required.

In \S2, we compare the rate of structure formation in
$\Lambda$CDM according to WMAP1 and WMAP3, on the scales relevant to
reionization. In \S3, we relate the history of reionization to the
growth of the mass fraction collapsed into source halos, and use this
to compare the reionization histories in WMAP3 and WMAP1 universes.
Our conclusions are summarized in \S4.

We adopt cosmological parameters
$(\Omega_mh^2$, $\Omega_bh^2$, $h$, $n_s$,
$\sigma_8)=(0.14,0.024,0.72,0.99,0.9)$ 
(Spergel et al. 2003) and $(0.127,0.022,0.73,0.95,0.74)$ 
(Spergel et al. 2006) for WMAP1 and WMAP3, respectively. The most
notable changes from old to new are: a reduction of 
normalization of the power spectrum on large scales ($\sigma_8=0.9
\rightarrow 0.74$) and more ``tilt'' ($n_s=0.99\rightarrow 0.95$). 
Throughout this paper, we use the transfer function of Eisenstein \&
Hu (1999).

\begin{figure}[t]
\plotone{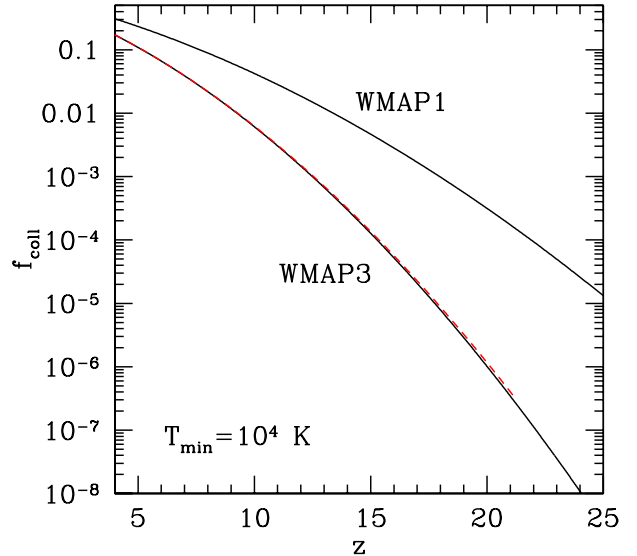}
\caption{Collapsed fraction vs. redshift for halos with virial temperatures
greater than $T_{\rm min}=10^4$ K, for WMAP1 and WMAP3, as labelled.  
Dashed curve, nearly on top of ``WMAP3'' curve, is ``WMAP1'' curve
with $1+z\rightarrow (1+z)/1.4$.
}
\vspace{-0.3cm}
\end{figure}

\section{Structure formation at high redshift}
A fundamental building block of models of reionization is the fraction
of the mass contained in virialized halos -- the ``collapsed
fraction'' -- the sites of ionizing photon production and
release. Using the Press-Schechter formalism (Press \& Schechter
1974), this collapsed fraction is given by: 
\begin{equation}
f_{\rm coll}(z)={\rm erfc}\left[\nu_{\rm min}(z)/\sqrt{2}\right],
\end{equation}
where $\nu_{\rm min}(z)\equiv \delta_c/[D(z)\sigma(M_{\rm min})]$,
$\sigma^2(M)$ is the variance in the present-day matter density field
according to linear perturbation theory, as filtered 
on the mass scale $M$, $D(z)$
is the linear growth factor ($D(z)\propto 1/(1+z)$ and 
$\delta_c=1.686$ in the matter-dominated era), and $M_{\rm min}(z)$ 
is the minimum mass for collapsed objects. For studies of reionization, 
the minimum mass is typically parameterized in terms of the minimum 
virial temperature, $T_{\rm min}$, of halos capable of hosting ionizing 
sources, $M_{\rm min}\simeq 4\times 10^7 M_\odot 
[(T_{\rm min}/{10^4 K})(10/(1+z))(1.22/\mu)]^{3/2},$\\
where $\mu=1.22$ for fully neutral gas (Iliev \& Shapiro 2001).

For $\Lambda$CDM, $\sigma(M)$ for
$M\sim 10^6-10^8 M_\odot$ is lower for WMAP3 than for WMAP1
by about 30 percent.  During reionization, when such halos 
are still rare, we expect their abundance to be exponentially
suppressed by this factor.  This is clearly shown in Figure 1, where the
new halo abundance and collapsed fraction are lower than the old ones
by 1-2 orders of magnitude. 
Since the threshold for halo collapse scales at these redshifts as 
$\delta_c/D(z)\propto 1+z$, structure 
formation on these mass scales is delayed by a factor $1/0.7 \sim 1.4$
in scale factor.  This is illustrated in Figure 2, where we plot 
$f_{\rm coll}(T>T_{\rm min}=10^4$ K) versus redshift and show that
the shift of 1.4 in scale factor provides an excellent description of
the delay in structure formation which results.

For the simplest possible reionization model, in which the universe is
instantly and fully ionized at some redshift $z_r$, the optical depth 
$\tau_{\rm es}\propto (1+z_r)^{3/2}$. If we assume that
$f_{\rm coll}(z_r)$ is a constant, so that reionization occurs when
the collapsed fraction reaches some threshold value, then our simple
estimate implies that the change in the cosmological parameters alone
reduces $\tau_{\rm es}$ by a factor of $1.4^{3/2}= 1.65$, from
$\tau_{\rm es}\sim 0.17$ to $\tau_{\rm es}\sim 0.1$.  In the next
section we discuss the motivation behind tying the reionization
history to the collapsed fraction. 

\section{Reionization History}

An important quantity in the theory of cosmic reionization is the
number of ionizing photons per hydrogen atom in the 
universe required to complete reionization\footnote{For simplicity, we
will neglect helium reionization.   
This does not effect our basic conclusions here.}. In the absence of 
recombinations, this ratio is unity. Given some observational constraint 
on the epoch of reionization, such as the onset of the GP effect
at $z\simeq 6.5$, we can
deduce that at least one ionizing photon per atom had to have been
released by that time. This 
ratio can then be used to predict other quantities, such as the
associated metal enrichment of the universe (e.g., Shapiro, Giroux, \& Babul 
1994) or the intensity of the near infrared background (e.g., Santos, 
Bromm, \& Kamionkowski 2002; Fernandez \& Komatsu 2006). Most 
models of cosmic reionization link the
ionized fraction of the IGM to the fraction of matter in collapsed
objects capable of hosting stars (e.g., Shapiro, Giroux, \& Babul
1994; Chiu \& Ostriker 2000; Wyithe \& Loeb 2003; Haiman \&
Holder 2003; Furlanetto, Zaldarriaga, \& Hernquist 2004; Iliev et
al. 2005; Alvarez et al. 2006).  For a given model, reionization is 
complete whenever the total number of ionizing photons emitted per 
hydrogen atom reaches some threshold value. Along with the escape 
fraction, star formation efficiency, and stellar initial mass function, 
the evolution of the collapsed fraction $f_{\rm coll}(z)$ forms the 
basis for calculation of this ratio and thus the reionization history.

To relate $\tau_{\rm es}$ to the halo abundance encoded
in $f_{\rm coll}$, it is necessary to determine the relationship
between the reionization history and the collapsed fraction.  If we
assume every H atom which ends up in a collapsed halo releases on 
average $f_\gamma(z)$ ionizing photons, and that $\epsilon_\gamma(z)$ 
is the number of ionizing photons consumed per ionized H atom, then we 
can write a simple relation between  $f_{\rm coll}$ and the mean ionized 
fraction,
\begin{equation}
x_e(z)=\frac{f_\gamma(z)}{\epsilon_\gamma(z)}f_{\rm coll}(z)\equiv
\zeta(z)f_{\rm coll}(z)
\label{xe}
\end{equation}
(e.g., Furlanetto, Zaldarriaga, \& Hernquist 2004).
For simplicity, we will assume a constant value,
$\zeta(z)=\zeta_0$ (this simplification does not affect our main
conclusions), and fix the value of $\zeta_0$ for a given  
$T_{\rm min}$, so that $\tau_{\rm es}=0.17$ for WMAP1. Reionization is
complete when the collapsed fraction 
reaches a threshold given by $f_{\rm coll}(z_r)\zeta_0=1$.

In Figure 3, we plot the value of $\nu_{\rm min}$ which
corresponds to $T_{\rm min}=10^4$~K.  As
mentioned in \S2, WMAP3 implies a
delay of structure formation by $\sim 1.4$ in scale factor. 
In the lower panel, we compare the reionization 
histories for WMAP1 and WMAP3 according to equation (\ref{xe}), for
the same efficiency $\zeta_0$. The same shift by a factor 1.4 in 
scale factor is also present in the reionization histories, which is not
surprising, since we have assumed that $x_e\propto f_{\rm coll}$,
and $f_{\rm coll}$ is a 
unique function of $\nu_{\rm min}$. As mentioned in \S2,
this change can account for a shift in the implied value of 
$\tau_{\rm es}$ from 0.17 to 0.1, quite close to the WMAP3 
value of $0.09\pm 0.03$. On the basis of this simple calculation, we conclude
that the reduction of $\tau_{\rm es}$ from WMAP1 to WMAP3 does not, itself,
significantly reduce the demand for high efficiency of ionizing
sources imposed previously by WMAP1.

\begin{figure}[t]
\plotone{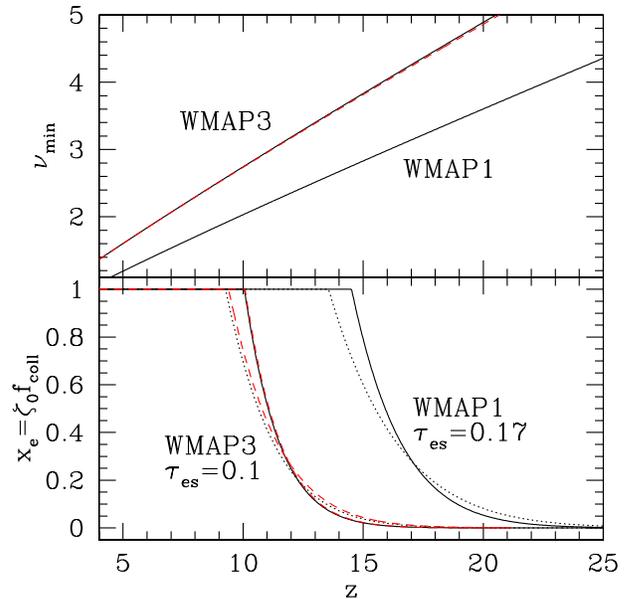}
\caption{
Evolution with redshift for WMAP1 and WMAP3, as labelled.
{\em Top:} Threshold for collapse, $\nu_{\rm min}$, for a halo with virial
temperature $10^4$~K.  Dashed curve, nearly on top of ``WMAP3'' curve,
is ``WMAP1'' curve with $1+z\rightarrow (1+z)/1.4$. 
{\em Bottom:} Reionization histories given by
  $x_e=\zeta_0f_{\rm coll}$, labelled by the corresponding values of
  $\tau_{\rm es}$, for $\zeta_0=170$ and 
  $T_{\rm min}=10^4$ K (solid), and $\zeta_0=35$ and 
  $T_{\rm min}=2\times 10^3$ K (dotted).  The two dashed curves are 
  ``WMAP1'' curves with $1+z\rightarrow (1+z)/1.4$.  
\vspace{-0.3cm}
        } 
\end{figure}

\subsection{Effect of recombinations}
Recombinations undoubtedly play an important role during
reionization. 
To first 
approximation, they should determine by what amount the parameter 
$\epsilon_\gamma(z)$ appearing in equation (\ref{xe}) exceeds unity. 
The quantity $\epsilon_\gamma-1$ is equal to the 
average number of recombinations that all ionized atoms must undergo
during reionization, 
$N_{\rm rec}$. As shown by Iliev et al. (2005), $N_{\rm rec}\simeq 0.6$ 
at percolation for large scale simulations of reionization that resolve 
all sources with masses greater than $\approx 2\times 10^9 M_\odot$, but 
do not resolve clumping of the IGM on scales smaller than $\approx 700$ 
comoving kpc. Surely, smaller scale structure affects reionization strongly 
(e.g., Iliev, Shapiro, \& Raga 2005; Iliev, Scannapieco, \& Shapiro 2005),
and therefore the number of recombinations per ionized atom is likely to be 
higher. For example, Alvarez, Bromm, \& Shapiro (2006) found that the
recombination time in the gas ionized by the end of the lifetime of a
100 $M_\odot$ star embedded in a $10^6 M_\odot$ halo at $z=20$ is 
$\sim 20$~Myr, roughly one tenth of the age of the universe
at that time.  

At the high redshifts considered here, the ratio of the age of the universe 
to the recombination time is proportional to $(1+z)^{3/2}$. 
Since structure formation is later for WMAP3 than for WMAP1 by a
factor of 1.4 in scale factor, photon consumption
due to recombinations is lower for WMAP3 by a factor $\sim1.4^{3/2}=1.65$. 
Even if recombinations dominate the consumption of ionizing photons
during reionization, therefore, 
the new {\em WMAP} data require an efficiency $\zeta_0$ which is 
at most a factor of only $\sim 1.65$ lower than that for
the first year data.  This is true even if clumping
increases toward lower redshift, since the evolution of clumping
follows structure formation and is, therefore, similarly delayed.
\vspace{0.5cm}

\section{Discussion}
We have shown that the new cosmological parameters reported for 
WMAP3 imply that structure formation at high redshift
on the scale of the sources responsible for reionization was delayed
relative to that implied by WMAP1. This
delay can account for the new value in $\tau_{\rm es}$ without
substantially changing the efficiency with which halos form stars.
Recombinations are fewer when reionization is later, but the reduction
is modest.  Even the IGM clumping factor on which this recombination
correction depends follows the delay in structure formation.

An important additional constraint on reionization is that it end at a
redshift $z\ga 6.5$, in order to explain the lack of a GP
trough in the spectra of quasars at $z\la 6.5$.  Because the GP
trough saturates at a very small neutral fraction, the quasar data
alone do not tell us when the universe became mostly ionized.  Indeed,
it is possible for the ionized fraction to have been quite
high already at high redshift $z\sim 10$ while there remained a
neutral fraction 
sufficiently high to satisfy the GP constraint at $z\sim 6.5$ (e.g.,
Choudhury \& Ferrara 2006).
While the universe may become mostly ionized well before $z\sim 6.5$,
it cannot be later than this, however.  Because of the shifting in
time of structure 
formation we have described, any model of reionization which previously
satisfied the WMAP1 $\tau_{\rm es}\sim 0.17$
constraint {\em and} became mostly ionized at $z\la 9$ would now
reionize too late to be compatible with the quasar observations.
Recently, Haiman \& Bryan (2006) used this fact to deduce that the
formation of massive Pop~III stars was suppressed in minihalos.   

While $\sigma_8=0.744$ and $n_s=0.951$ are the most likely values obtained
from the new WMAP data alone, there remain significant uncertainties.
When combined with other data sets such as large-scale structure
(e.g., Spergel et al. 2006) and the Lyman-$\alpha$ forest
(e.g., Viel, Haehnelt, \& Lewis 2006; Seljak, Slosar, \& McDonald
2006), important differences arise.  While these differences
may seem small from the point of view of the statistical error of the
observational data, the implications for reionization can be quite
dramatic, as we have seen here. It is also important to note that
these measurements of the power spectrum are on scales much larger
than those relevant to the sources of reionization. As such, the
theory of reionization requires us to extrapolate the power spectrum
by orders of magnitude beyond where it is currently measured.  This
means that the study of reionization is crucial to extending the
observational constraints on the origin of primordial density
fluctuations (e.g., by inflation) over the widest range of wavenumbers
accessible to measurement.  Direct 
observations of the high redshift universe such as 21-cm tomography
(e.g., Iliev et al. 2002; Zaldarriaga, Furlanetto, \& Hernquist 2004; 
Shapiro et al. 2006; Mellema et al. 2006) and large-aperture infrared
telescopes such as JWST promise to diminish the uncertainties which
currently prevent us from making reliable statements about the nature
of the first sources of ionizing radiation. 

\acknowledgments
We are grateful for the support of NASA Astrophysical Theory Program
grants NAG5-10825 and NNG04G177G, a DOE Computational Science Graduate
Fellowship to M.~A.~A., and discussion with Volker Bromm, Leonid
Chuzhoy, and Eiichiro Komatsu.

\end{document}